\documentstyle[aps]{revtex} \input{psfig.sty}
\begin{document}
\setlength{\topmargin}{5pt}
\draft

\title{Evaporative cooling of trapped fermionic atoms}
\author{W. Geist, A. Idrizbegovic, M. Marinescu, T. A. B. Kennedy, and L. You}
\address{School of Physics, Georgia Institute of Technology,
Atlanta, GA 30332-0430}
\date{\today}
\maketitle

\begin{abstract}
We propose an efficient mechanism for the evaporative cooling of
trapped fermions directly into quantum degeneracy. Our idea is
based on an electric field induced elastic interaction between
trapped atoms in spin symmetric states. We discuss some novel
general features of fermionic evaporative cooling and present
numerical studies demonstrating the feasibility for the cooling of
alkali metal fermionic species $^6$Li, $^{40}$K, and
$^{82,84,86}$Rb.  We also discuss the sympathetic cooling of
fermionic hyperfine spin mixtures, including the effects of
anisotropic interactions.
\end{abstract}


\narrowtext

\section{Introduction}
The success of atomic Bose-Einstein condensation (BEC) has opened
a major research area \cite{bec,mit-1,bec-2} in weakly interacting
quantum gases \cite{edwards}. Experimental breakthroughs have been
achieved thanks to several remarkable advances in atomic, molecular,
and optical physics. In particular, the
laser cooling and trapping of neutral particles \cite{ct,nobel},
and the ingenious application of the evaporative cooling
technique \cite{doyle,druten,walraven}.

Evaporative cooling was first developed for magnetically confined
atomic H \cite{doyle}, and is based on a simple principle seen in
everyday life \cite{druten,walraven}. In a quasi equilibrium system
the particles most likely to evaporate are the most energetic. In
doing so, the particle leaves behind a system with a lower average
energy per particle, i.e., a cooled system. Less appreciated
perhaps, is that forced evaporative cooling is the key to reaching
quantum degeneracy in alkali metal systems. A sufficient condition
for successful evaporative cooling is the threshold ``run-away''
requirement, when an increased collision rate is achieved despite
the loss of atoms. This technique was adapted to alkali atomic
systems in the first generation of BEC experiments \cite{alkali}.
It has since been demonstrated to work over a wide range of
temperatures and densities, and has been applied to all successful
BEC experiments this far \cite{edwards}.

With such a powerful technique one may ask why it has not yet been
used
 to create degenerate atomic Fermi gases ?
A careful examination of current magnetic trapping set-ups helps to
clarify the fundamental problem. Alkali metal atoms
have two degenerate manifolds of hyperfine ground states
$f=I\pm 1/2$, due to the coupling between the nuclear $\vec I$ and
electronic spin $\vec s$ $(s = 1/2)$.
A typical magnetic trap arrangement confines
 the low field seeking Zeeman state
with maximum spin polarization $m_f=f$ (at the moment we assume a
single $m_f$ value is trapped, mixtures of two different $m_f$ values will
be discussed later). Eigenstates of $\vec F ^2$, where
$\vec F=\vec f+\vec f$ is the total hyperfine spin of two colliding atoms,
 have definite spin exchange symmetry
given by $(-1)^{2f-F}= +1$. Thus only the
partial waves for which the relative orbital angular momentum $l$
is odd contribute to the antisymmetrized scattering amplitude.
From
low energy atomic collision theory, we know that the partial wave
phase shifts are $\delta_l(k) \propto k^{2l+1}$ for a short ranged
potential, typical of alkali metal atoms in their ground state.
Therefore only the s-wave ($l=0$) has a non-zero contribution in
the low energy $(k\to 0$) limit, characterized by a scattering
length $a_{\rm sc}=-\lim_{k\to 0}\delta_0(k)/k$. One then obtains
the elastic cross section $\sigma_B=8\pi a_{\rm sc}^2$ for bosons,
while for fermions $\sigma_F \rightarrow 0$, in the limit $k \rightarrow 0$.
 Without the thermal equilibration produced by
elastic collisions, evaporative cooling can not be directly applied
to a single fermionic species.
This is the same fundamental reason that BCS states
for trapped fermionic gases are not easily accessible
\cite{Stoof,Kagan}. Obviously, if two or more spin components are
simultaneously trapped, spin-statistics does not prohibit s-wave
collisions between the spins at low energy, so that sympathetic cooling
is in principle possible.

In this paper we discuss an efficient mechanism for the evaporative
cooling of trapped fermions. The mechanism employs an external
electric field to polarize atoms, enabling cooling at low energy.
We also discuss the sympathetic cooling of fermionic mixtures. The
paper is organized as follows. In section II we begin with a
discussion of the major results for collisions between atoms in an
external electric field \cite{mm}. We then outline our proposal for
the evaporation kinetics of fermions, closely following the
approach developed in \cite{luiten,kris} for bosons. We present our
numerical studies and estimate the time scales for fermionic
species $^6$Li and $^{40}$K. In section III we discuss sympathetic
cooling of fermionic spin mixtures, along the lines of the Jila
experiment \cite{jin}. Finally, in section IV we present our
conclusions.

\section{ Collision kinetics of atoms in an electric field}
As shown in detail in ref. \cite{mm}, polarized atoms in
an external electric field interact through the
dipole interaction
\begin{equation}
V_E(R)=-\frac{C_E}{R^3} P_2(\cos\theta),
\label{a2}
\end{equation}
in addition to the usual interatomic potentials. Here,
$C_E=2E^2\alpha_1^A(0)\alpha_1^{B}(0)$ is the
electric field induced dipole interaction coefficient and $\alpha_1^{A,B}(0)$
are the static atomic dipole polarizabilities of the two atoms
denoted by $A$ and $B$, respectively. The angle between the directions
of the electric field ( of amplitude $E$ )
and the internuclear axis is denoted by $\theta$, and $P_2$ is the
 Legendre polynomial of order 2. The presence of this anisotropic interaction
term completely changes the low energy collision physics.
In particular {\it all} partial waves now contribute
to the low energy scattering cross sections since one
can prove analytically that $\delta_l(k)\sim k$ for all $l$ \cite{mm}.
The unsymmetrized low energy scattering amplitude is
\begin{equation}
F({\vec k},{\hat R})={4\pi}\sum_{lm,l'm'}
t_{lm}^{l'm'}(k)Y_{l'm'}^*({\hat k})Y_{lm}({\hat R}),
\label{ca}
\end{equation}
where $t_{lm}^{l'm'}$ are the reduced T-matrix elements, the multi-channel
analogues of the scattering length. Here the incident relative
momentum of the two colliding atoms is denoted by $\vec k = k \hat k$,
and $\vec k'= k \hat R  $ is
the relative momentum after the collision.

For a gas of atoms in an electic field we can now derive the collision
integral for the
quantum Boltzman equation (QBE). The QBE is given by
\begin{eqnarray}
\left(\frac {\partial} {\partial t} + \frac {\vec p}{m} \cdot
\nabla_{\vec r}-\nabla_{\vec r} U(\vec r) \cdot \nabla_{\vec p}\right)
\, f(\vec r,\vec p,t) = {\cal I}(\vec r,\vec p,t),
\label{pkinetic}
\end{eqnarray}
where $f(\vec r,\vec p,t)$ is the phase space distribution function, and
 $U(\vec r)$ the external trapping potential. Using Eq. (\ref{ca})
we arrive at the following collision integral
\begin{eqnarray}\label{kernal}
{\cal I}(\vec r,\vec p_4,t) &=&
\frac{2}{m (2\pi\hbar)^3} \int d \vec p_3 d\hat q\,'
\left|F_{B/F}
\left({1\over 2}(\vec p_3-\vec p_4),{1\over 2}(\vec p_1 - \vec p_2)\right)\right|^2
\left|\frac {\vec p_3-\vec p_4}{2}\right|\nonumber \\
&&\Big [ f(\vec r,\vec p_1) f(\vec r,\vec p_2)
[1+\eta f(\vec r,\vec p_3)][1+\eta f(\vec r,\vec p_4)]-
[1+\eta f(\vec r,\vec p_1)] [1+\eta f(\vec r,\vec p_2)]
f(\vec r,\vec p_3)f(\vec r,\vec p_4)\Big],
\end{eqnarray}
where $F_{\phi}$ is the symmetrized, unsymmetrized, or the anti-symmetrized
scattering amplitude for bosonic ($\eta=1$), classical ($\eta=0$), or
fermionic atoms ($\eta=-1$), respectively.
We have used the following notation. Two atoms with
incoming momentum $\vec p_3$ and $\vec p_4$ collide and
produce an outgoing pair with momentum $\vec p_1$ and $\vec p_2$.
 The scattering amplitude from
an initial relative momentum
$\vec q=(\vec p_3-\vec p_4)/2$ to a final
momentum $\vec q\,'=(\vec p_1-\vec p_2)/2$
is given by $F(\vec q,\vec q\,')$.
For elastic collisions one has $q=q\,'$ in addition to the conservation
of the total momentum $\vec P=\vec p_3+\vec p_4=\vec p_1+\vec p_2$
[assuming the collision to be localized in space at a fixed $U(\vec r)$]
and energy $\epsilon_3+\epsilon_4=\epsilon_1+\epsilon_2$.
The differential cross section
for scattering into the (relative) solid angle $d\hat q\,' $ is given by
$d \sigma = |F(\vec q, \vec q\,')|^2 d \hat q\,'$,
and $|F(\vec q, \vec q\,')|^2 = |F(\hat q,\hat q\,')|^2$ for the
low energy collisions under consideration. The approximation of
localized collisions within a small spatial region compared with the
variations of $U(\vec r)$ allows us to neglect the $\vec P$
dependence of $F$.

Assuming an ergodic distribution for the system, we introduce the
energy dependent single particle distribution function
$n(\epsilon,t)$ and obtain the kinetic equation in ergodic
approximation \cite{luiten},
\begin{eqnarray}
\rho(\epsilon_4){\partial\over \partial t} n(\epsilon_4,t)
&\approx & \frac {m\sigma_{\phi}}{\pi^2 \hbar^3}
\int d \epsilon_{1}\,d \epsilon_{2}\,d \epsilon_{3}\,
\delta(\epsilon_{1}+\epsilon_{2}-\epsilon_{3}-\epsilon_{4})
\rho(\min [\epsilon_{1},\epsilon_{2},\epsilon_{3},\epsilon_{4}])\nonumber\\
&&\Big [n(\epsilon_1)n(\epsilon_2)
[1+\eta n(\epsilon_3)][1+\eta n(\epsilon_4)]-
[1+\eta n(\epsilon_1)][1+\eta n(\epsilon_2)]
n(\epsilon_3)n(\epsilon_4)\Big],
\label{ekinetic}
\end{eqnarray}
where $\rho(\epsilon)$ is the density of states.
Our result is formally similar to the standard result
for isotropic s-wave collisions in the absence of an electric field
\cite{luiten,kris}, when the s-wave collision cross section is replaced by the
effective scattering cross sections $\sigma_{\phi}$. In reaching the
form of Eq. (\ref{ekinetic}),
the collision integral has been approximated by assuming that the fast
varying part in the $\cos\theta_{\vec q}$ and $\cos\theta_{\vec q\,'}$
integration comes from
the delta functions $\delta[\epsilon-p_i^2/2m-U(r)]$
rather than the scattering amplitude $F_{\phi}$,
thus we have extracted the scattering cross section
factor out of the integrand, and replaced it by its average
\begin{eqnarray}
\int\! d\hat q\,'d \hat q{1\over 4\pi}|F(\hat q,\hat q')|^2&=&
2\times 4\pi\sum_{lm,l'm'}^{l,l' {\rm even/odd}}|t_{lm}^{l'm'}|^2
\equiv \sigma_{\eta},
\end{eqnarray}
where the additional factor of $2$ occurs in the Bose/Fermi case
due to the symmetrization/anti-symmetrization.
A specific discussion of p-wave scattering is
presented later.

For highly confined atoms inside magnetic traps, corrections due to
discrete level structure need to be included. When a long range
type of potential (\ref{a2}) is involved, one can not in general
approximate the atom-atom interaction potential by a contact
$\delta(\vec r -\vec r')$ form as is usually adopted in the s-wave
collision models. Special care is needed to find the scattering
matrix element for atoms in different trap states. The quantum
kinetic equation now takes the form (for bosons and fermions)
\begin{eqnarray}
\frac{dn_{{\vec i}}}{dt}&=&
\sum_{{\vec j,\vec k,\vec l}}
\delta_{\epsilon_{{\vec i}} +\epsilon_{{\vec j}}, \epsilon_{{\vec k}}
+\epsilon_{{\vec l}}}\gamma_{\vec i,\vec j;\vec k,\vec l}
\nonumber\\
&&\left( n_{{\vec k}} n_{{\vec l}} (1\pm n_{{\vec i}})
(1\pm n_{{\vec j}})-
n_{{\vec i}} n_{{\vec j}} (1\pm n_{{\vec l}})
(1\pm n_{{\vec k}})\right),
\label{qkinetic}
\end{eqnarray}
where $n_{{\vec i}}=n(\epsilon_{\vec i})$ is the population in
level $\epsilon_{\vec i}=\hbar (i_x\omega_x+i_y\omega_y+i_z\omega_z)$
with $\omega_x$, $\omega_y$, and $\omega_z$ the trap frequencies in
the $x,\;y,\;$ and $z$ directions.

For s-wave collisions, when the contact potential is valid,
the collisional matrix element is given by
\begin{eqnarray}
\gamma_{\vec i,\vec j;\vec k,\vec l}\propto
\left|\int \Phi_{\vec i}^*(\vec r) \Phi_{\vec j}^*(\vec r)
\Phi_{\vec k}(\vec r) \Phi_{\vec l}(\vec r) d \vec r\right|^2,
\label{gs}
\end{eqnarray}
where $\Phi_{\vec i}(\vec r)$ denotes the wave function for trap state
$\vec i = (i_x,i_y,i_z)$.
For the anisotropic collisions discussed here we have
\begin{eqnarray}
\gamma_{\vec i,\vec j;\vec k,\vec l}={2\pi\over\hbar}
{1\over \hbar\omega} \left|4\pi
\sum_{lm,l'm'}t_{lm}^{l'm'}\int d \vec p\, I_{\vec i,\vec j}^{l'm'*}(\vec p)
I_{\vec k,\vec l}^{lm}(\vec p)\right|^2,
\label{ggs}
\end{eqnarray}
with
\begin{eqnarray}
I_{\vec i,\vec j}^{l'm'}(\vec p)=\int d \vec q\,'
\Phi_{\vec i}({\vec p/2}+\vec q\,')
\Phi_{\vec j}({\vec p/2}-\vec q\,')
Y_{l'm'}^{*}(\hat q\,'),
\end{eqnarray}
where $\Phi_{\vec i}(\vec p)$ is the wavefunction in momentum space.

Clearly, apart from final state quantum
statistical factors, the collision kernel
on the rhs of Eqs. (\ref{ekinetic}) and (\ref{qkinetic})
takes the same form as that for classical atoms with s-wave collisions.
For fermions, the effective cross section $\sigma_{-}$ provides
an opportunity for investigation of the ergodic
kinetics as has been done previously for bosons \cite{murray,zoller}.
Therefore the net effect of the electric field is to polarize
atoms, modify their collisional properties and thereby induce
a non-zero elastic cross section at low energy.
 The electric field dependence of the
effective cross sections $\sigma=\sigma_{\pm}$ allows one to control
the time scale of evaporation by adjusting the
strength of the electric field $E$.
In addition, since the microscopic collision processes for
polarized atoms are anisotropic, kinetic motions in different
spatial directions are mixed. The validity of
the ergodic approximation is thus reinforced.

\subsection{Results}
In this subsection we report our numerical studies of
fermionic evaporative cooling in an electric field.
 We focus primarily on the qualitative differences between fermions
and bosons in this paper, and the fundamental issues
relating to fermionic evaporation. We will approximate
the scattering matrix element $\gamma_{\vec i,\vec j;\vec k,\vec l}$
by the equivalent s-wave contribution, retaining only the leading
$t^{00}_{00}$ contribution in (\ref{ggs}).
We believe that this approximation should not produce any qualitative
differences from the full model in the results presented.

\begin{figure}
\psfig{figure=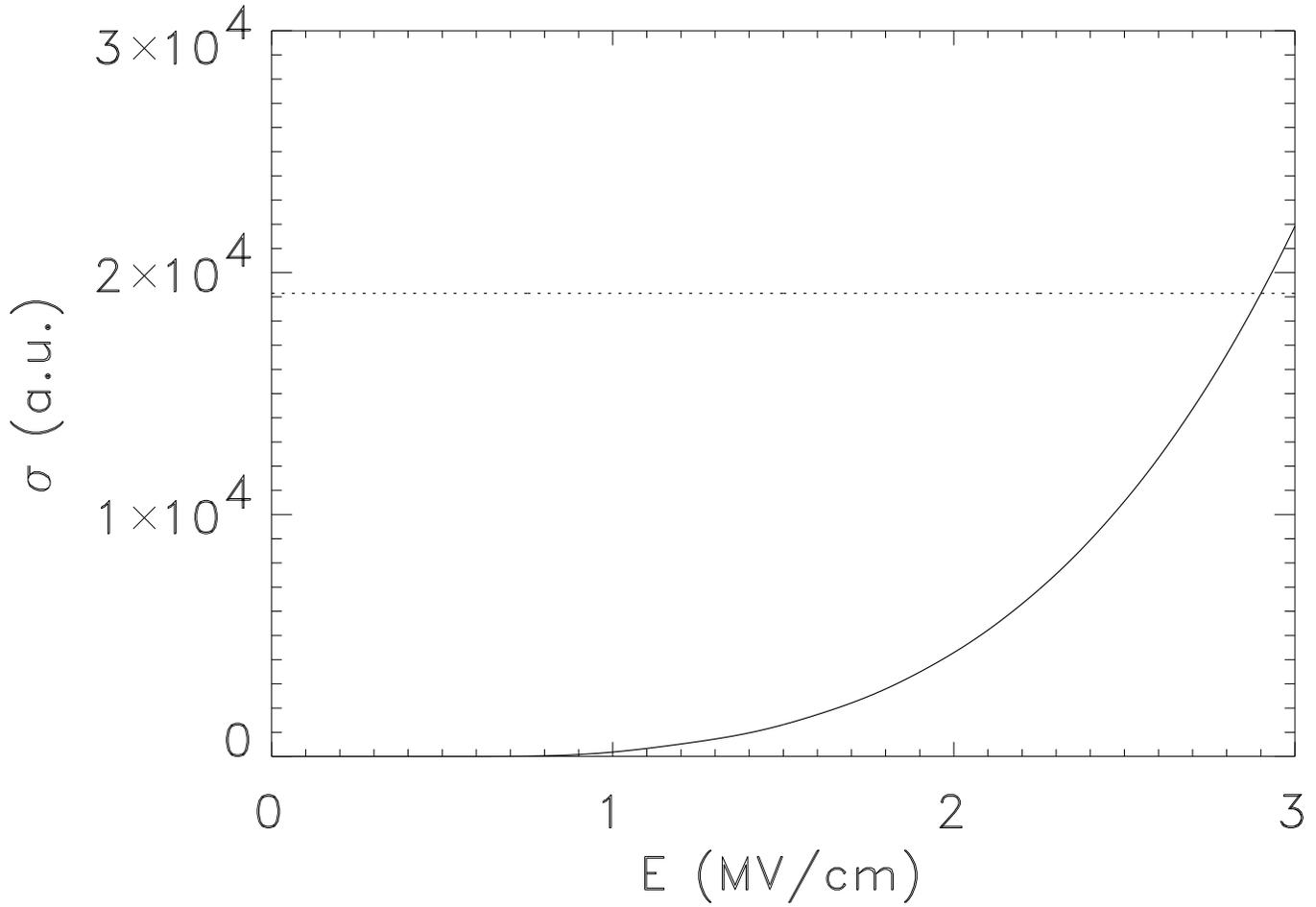}
\vspace{0.5in}
\caption{The low energy collision cross section for spin triplet $^6$Li
(fermion). The dashed horizontal line indicates the field free
value of triplet $^7$Li [boson, $a_{\rm sc}=-27.6$ (a.u.)].
}
\label{fig1}
\end{figure}

\begin{figure}
\psfig{figure=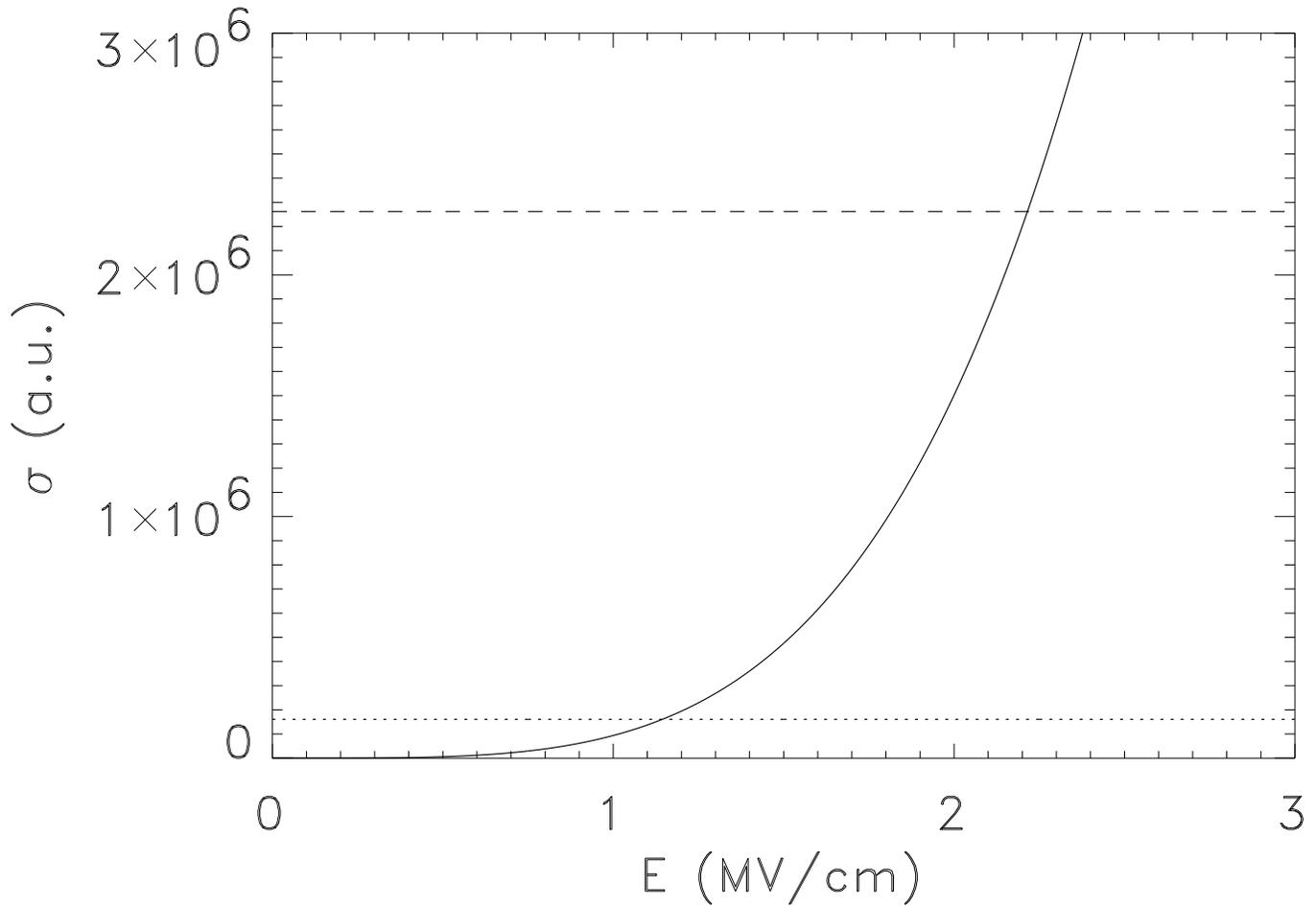}
\vspace{0.5in}
\caption{The same as in Fig. 1 but for $^{40}$K (fermion).
The dotted line is the field-free reference value for $^{39}$K
[boson $a_{\rm sc}\sim 80$ (a.u.)] and the dashed line the
field-free reference value for $^{41}$K [boson $a_{\rm sc}\sim 300$
(a.u.)].}
\label{fig2}
\end{figure}

\begin{figure}
\psfig{figure=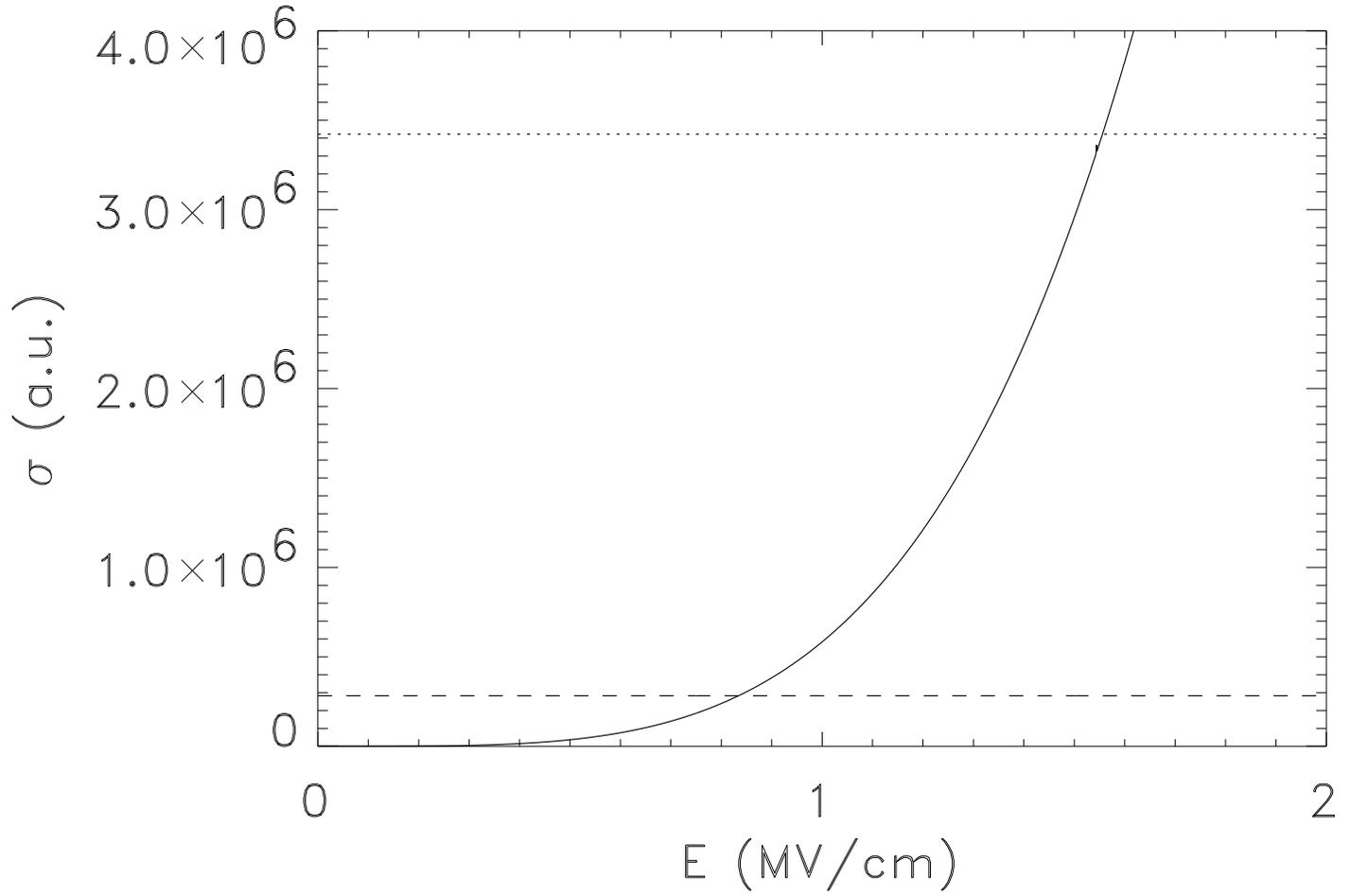}
\vspace{0.5in}
\caption{The same as in Fig. 1 but for $^{86}$Rb (fermion)
which is similar to the results of $^{82}$Rb and $^{84}$Rb.
The dotted line is the field-free reference value
for $^{85}$Rb [boson, $a_{\rm sc}\sim 369$ (a.u.)]
and the dashed line the field-free reference value
for $^{87}$Rb [boson, $a_{\rm sc}\sim 106$ (a.u.)].
}
\label{fig3}
\end{figure}

In the early stages of evaporative cooling, when the effect of
phase space blocking term $[1-f(\vec r,\vec p)]$
or $(1-n_\epsilon)$ is unimportant, the kinetic term reduces
precisely to that of a classical gas with an
effective cross section $\sigma_{-}$. The evaporative
kinetics of such a case have been extensively studied
\cite{druten,walraven,luiten,kris,murray,zoller}. The classical
run-away conditions can be achieved provided loss mechanisms
are controlled. For $^6$Li, $^{40}$K, and $^{82,84,86}$Rb, we have performed
detailed multichannel collision calculations for atoms in an electric field.
 For fields of the order of a few (MV/cm) we find that one can readily
induce an effective scattering cross section as large as for
their bosonic counterparts $^7$Li, $^{39}$K, $^{41}$K,
 $^{85}$Rb, and $^{87}$Rb.
Our results are displayed in Figs. \ref{fig1}, \ref{fig2}, and
\ref{fig3}. The case of $^6$Li is particularly interesting, since
the triplet scattering cross section reaches the field-free value
of bosonic $^7$Li ($8\pi a_{T}^2$) with $a_{T}=-27.6$\,(a.u.) at a
value of $E\sim3$ (MV/cm) \cite{bec-2}. For $^{40}$K, it reaches
the field free bosonic isotope value of $a_{T}\sim 80$ (a.u.) for
$^{39}$K and $a_{T}\sim 300$ (a.u.) \cite{cote} for $^{41}$K at
values of $E\sim 2 $ (MV/cm) and $E\sim 1 $ (MV/cm), respectively.
For $^{82,84,86}$Rb, it reaches the field free bosonic isotope
value of $a_{T}\sim 369$ (a.u.) for $^{85}$Rb and $a_{T}\sim 106$
(a.u.) for $^{87}$Rb \cite{chris} at values of $E\sim 0.8 $ (MV/cm)
and $E\sim 1.5 $ (MV/cm), respectively.

The promising scenario discussed above can not be simply
extrapolated to the regime of quantum degeneracy. The effect of
phase space blocking significantly reduces the collision rates at
quantum degeneracy. Time scales for evaporation are greatly
prolonged for fermions, while for bosons the final state
stimulation factor $(1+n_\epsilon)$ further enhances the classical
run-away evaporation. Of course this type of slowing is not unique
to our scheme of fermionic evaporation, it is a generic feature of
fermionic quantum statistics. As such, it is also present for other
fermionic cooling schemes such as sympathetic cooling with bosons.
One should realize that in this regime of phase space density the
energy scale involved is already extremely low, and the crucial
challenge is in keeping the inelastic collision and loss rates
down.

\begin{figure}
\psfig{figure=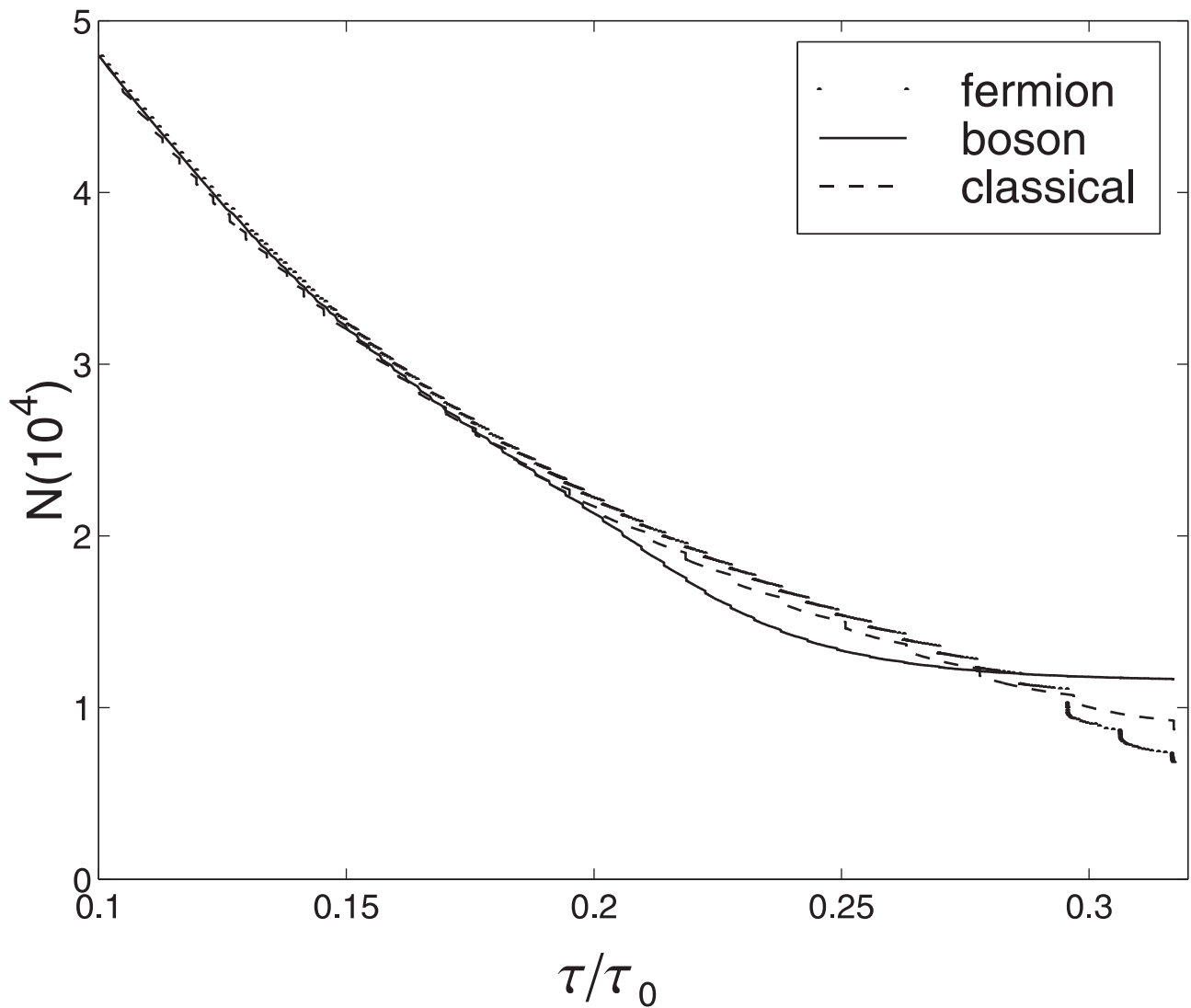}
\vspace{0.5in}
\caption{Comparisons of the evaporative cooling for
bosons, fermions, and classical atoms. The remaining number of
atoms in the trap. The steps in the number of fermions for larger
times is due to cutting discrete energies for a distribution
function which does not rethermalize fast enough.}
\label{fig4}
\end{figure}

\begin{figure}
\psfig{figure=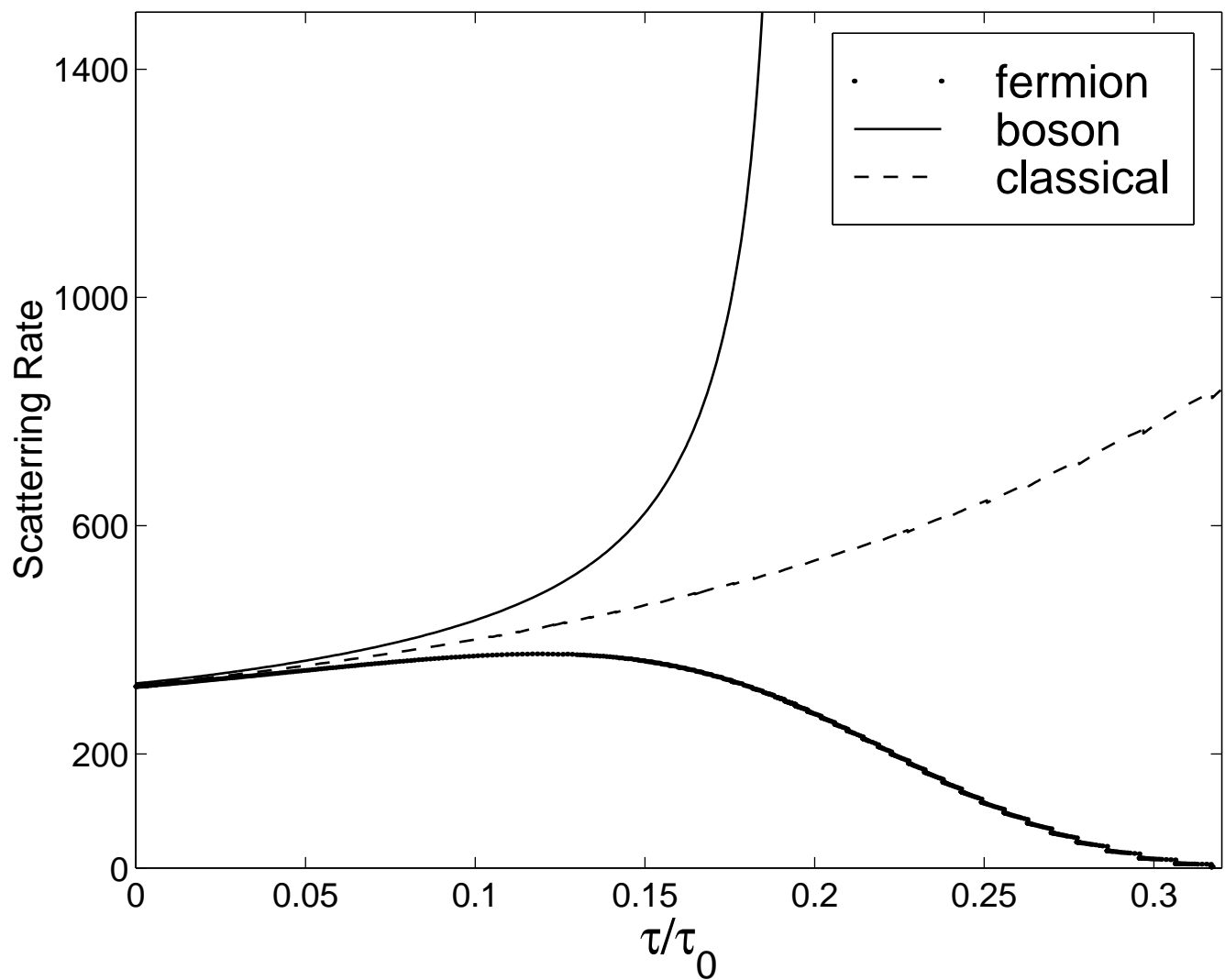}
\vspace{0.5in}
\caption{The same as in Fig. 4, but for the numerically
computed collision rate.
}
\label{fig5}
\end{figure}

\begin{figure}
\psfig{figure=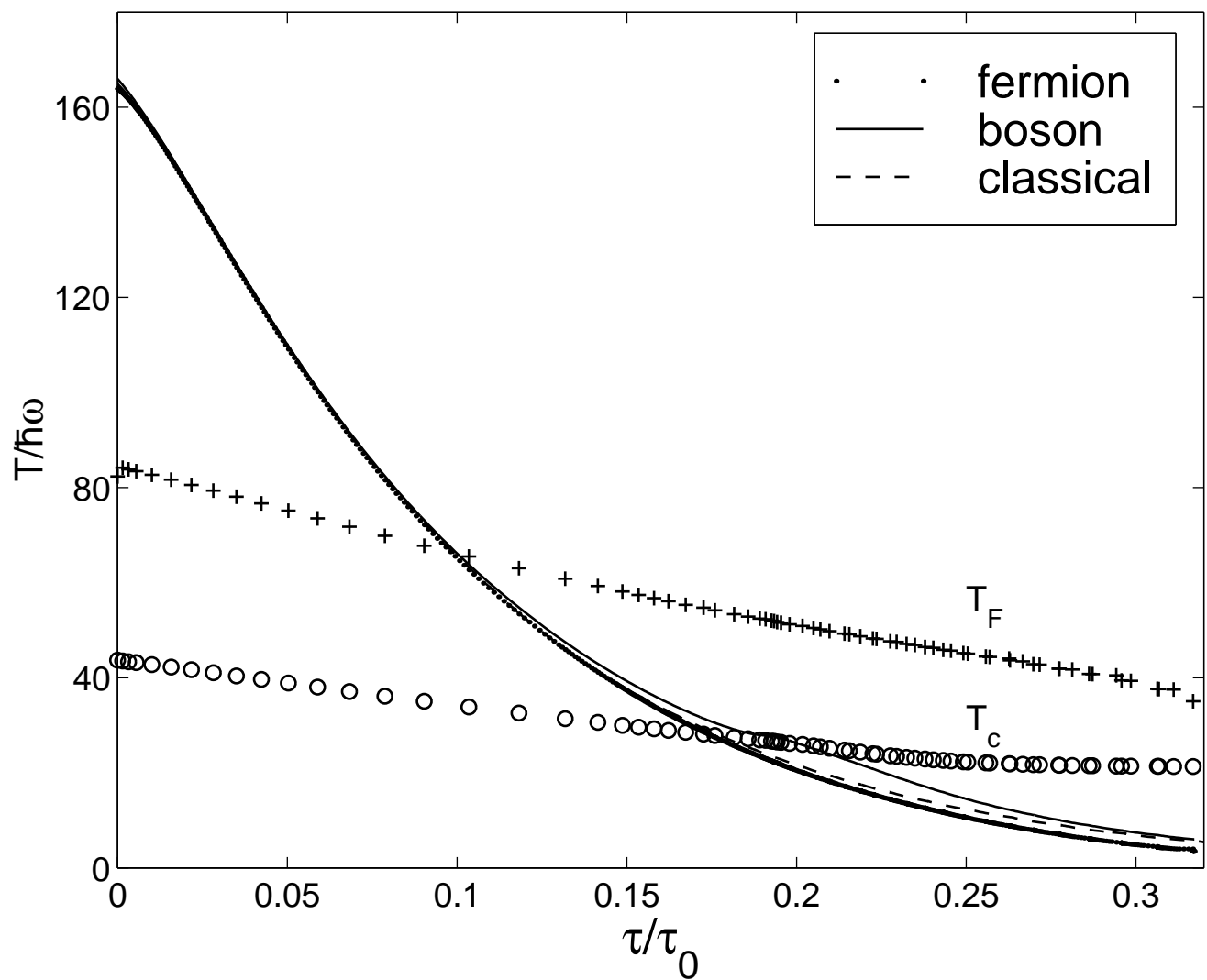}
\vspace{0.5in}
\caption{The same as in Fig. 4, but for the fitted temperatures
of the remaining atomic distributions.
}
\label{fig6}
\end{figure}

In numerical studies with Eq. (\ref{qkinetic}) for a variety for
parameters, we have found that down to the limit of a trapped gas
temperature of the order of the Fermi energy $\mu_F$ (typically
$\sim 0.5\mu_F$), the evaporation proceeds as if the atoms were
classical or bosonic, i.e., the final state blocking effect is
unimportant. In Figs. \ref{fig4} and \ref{fig5} we display
comparisons of evaporative cooling for bosons, fermions, and
classical atoms (with the same collision cross section $\sigma$)
contained in an isotropic harmonic trap with trap frequency
$\hbar\omega$. The system contains $N=10^5$ atoms initially, which
are assumed to obey the Maxwell Boltzmann distributions with an
initial temperature $T=164.92\,(\hbar\omega)=2T_F$, i.e., twice the
initial Fermi temperature. The evaporation is effected by ramping
down the trapping potential with a cut $\epsilon_{\rm cut}(t)
=\epsilon_0 e^{-\gamma_{\rm cut} t}$. The time scale of the
evaporation is $\tau=t/\tau_0$ with $1/ \tau_0\equiv (m\sigma /
\pi^2\hbar^3) (\hbar\omega)^2$. The evaporation rate
is $\gamma_{\rm cut}=\gamma_{\rm coll}(0)/30$, with $\gamma_{\rm
coll}(0)$ the initial collision rate. The time dependent collision
rate is given by
\begin{eqnarray}
\gamma_{\rm coll}(t)&=&\frac{1}{N(t)}
\sum_{\epsilon_{\vec i},\epsilon_{\vec j},\epsilon_{\vec k},\epsilon_{\vec l}}
\delta_{\epsilon_{\vec i}+\epsilon_{\vec j},\epsilon_{\vec k}+\epsilon_{\vec l}}
\gamma (\epsilon_{\vec i},\epsilon_{\vec j},\epsilon_{\vec k},\epsilon_{\vec l})
\nonumber\\
&&\Big([n_{\epsilon_{\vec k}}(t)n_{\epsilon_{\vec l}}(t)
[1+\eta n_{\epsilon_{\vec j}}(t)]
[1+\eta n_{\epsilon_{\vec i}}(t)]\nonumber\\
&&-n_{\epsilon_{\vec i}}(t) n_{\epsilon_{\vec j}}(t)
[1+\eta n_{\epsilon_{\vec k}}(t)][1+\eta n_{\epsilon_{\vec l}}(t)]\Big).
\end{eqnarray}
In Fig. \ref{fig6} the temperature $T(t)$ is determined by fitting
a Maxwell, Bose, or Fermi distribution function with a temperature
$T(t)$ and fugacity ${\sf z}$ to the total number of remaining
atoms $N(t)=\sum_{\epsilon}\rho(\epsilon)n_\epsilon[{\sf z}(t),T(t)]$
and the average energy
$\bar\epsilon=\sum_{\epsilon}\rho(\epsilon)\epsilon\,
n_\epsilon[{\sf z}(t),T(t)]$. In the trap, the Fermi temperature
$T_F$ and the critical temperature for Bose-Einstein condensation
$T_C$ are given by $k_B T_F = \hbar \omega (6 N)^{1/3}$
and $k_B T_C= \hbar \omega (N/1.202)^{1/3}$.

\section{Sympathetic cooling of hyperfine fermionic mixtures}
In a recent experiment the JILA group has succeeded in implementing
rf evaporation on a trapped fermionic cloud containing two spin
states \cite{jin}. The effect of the low energy p-wave suppression
was clearly demonstrated in their measurements of the collision
rates. Our general analysis of anisotropic collisions in this paper
can be adapted to an analysis of this experiment, in particular, by
specializing to the case of s-wave collisions between the two
trapped spin states, and p-wave collisions for each of the two spin
states. We first discuss the appropriate collision integrals again
assuming that trap confinement is not important during the
collisions, i.e., when the ground state size of the trap is much
larger than the effective range of the interatomic potential; this
is true for almost all magnetic traps currently employed in
laboratories. The scattering cross section
\begin{eqnarray}\label{inv}
&&|F_{B/F}(\vec q, \vec q')|^2 d\hat q'\nonumber\\
&&=\left(\frac{d\sigma(\vec q, \vec q')}{d \hat q'}\right)_{\rm lab}
d \hat q'_{\rm lab}
=\frac{d\sigma(\vec q, \vec q')}{d \hat q'}d \hat q'
\end{eqnarray}
is invariant in different reference frames, and this enables us to
 write
the differential scattering cross section for p-wave
scattering as
\begin{eqnarray}
\frac{d\sigma_p}{d \hat q'}=
\alpha q^4 \cos^2\theta_{\hat q}\cos^2 \theta_{\hat q'},
\end{eqnarray}
where $\theta,\theta'$ specify the directions of the relative
momentum before and after the scattering. The constant $\alpha$ is
obtained from the experimental data \cite{jin} in terms of an
equivalent s-wave scattering cross-section
$\sigma_p(T_0)=\sigma_s$ at a temperature $T_0$, i.e.,
\begin{eqnarray}
\alpha \overline {q^4} (4\pi)=\sigma_s,
\end{eqnarray}
where $\overline {q^4}$ is the average of $q^4$ taken with respect
to a Boltzmann distribution. This yields
\begin{eqnarray}
\alpha={\sigma_s\over 15\pi} {1\over (mk_BT_0)^2}.
\end{eqnarray}
We can then substitute Eq. (\ref{inv})
into (\ref{kernal}), and integrate over the
phase space $d\vec rd\vec p_4$ again using
the assumption of ergodicity
\begin{eqnarray}
f(\vec r,\vec p)=\int d\epsilon\,
n(\epsilon)\,\delta[\epsilon-p^2/2m-U(r)],
\end{eqnarray}
to obtain the ergodic Boltzmann equation for p-wave
part of the scattering
\begin{eqnarray}\label{pwave}
\rho(\epsilon_4){\partial\over \partial t} n(\epsilon_4,t)
&= & \frac {m\sigma_{s}}{\pi^2 \hbar^3}\frac{1}{45(k_BT_0)^2}
\int d \epsilon_{1}\,d \epsilon_{2}\,d \epsilon_{3}\,
\delta(\epsilon_{1}+\epsilon_{2}-\epsilon_{3}-\epsilon_{4})
\nonumber\\
&&\rho(\epsilon_<)\frac{3 \epsilon_>-\epsilon_<}
{(\epsilon_>-\epsilon_<)^3}
(\epsilon_1-\epsilon_2)^2(\epsilon_3-\epsilon_4)^2
\nonumber\\
&&\Big[n(\epsilon_1)n(\epsilon_2) [1+\eta n(\epsilon_3)][1+\eta
n(\epsilon_4)]- [1+\eta n(\epsilon_1)][1+\eta n(\epsilon_2)]
n(\epsilon_3)n(\epsilon_4)\Big],
\end{eqnarray}
where $\epsilon_>$ ($\epsilon_<$) is the largest
(smallest) of the four energies $\epsilon_i$.
We see that the collisions become ineffective as soon
as the energy scales become less than $k_BT_0$.
To obtain a classical collision rate we substitute
the Maxwell Boltzmann distribution
\begin{eqnarray}
n(\epsilon)={\sf z}^{-1}e^{-\epsilon/k_BT},
\end{eqnarray}
into equation [\ref{pwave}], set $\eta=0$, and replace the sums
by integrals and integrate the first term on the r.h.s. over
$\epsilon_4$. This yields the p-wave collision rate as
\begin{eqnarray}
\gamma_{p}&=& N(t)
\frac {m\sigma_{s}}{\pi^2 \hbar^3}(\hbar\omega)^2
{I_p\overline T \over 45\overline T_0^2 } ,
\end{eqnarray}
with the scaled temperatures $\overline T=k_BT/\hbar\omega$ and
$I_p=5$ is the value of the fourfold integral
\begin{eqnarray}\label{ip}
I_p&=&\int dx_{1}\,dx_{2}\,dx_{3}\,dx_{4}
\delta(x_{1}+x_{2}-x_{3}-x_{4})e^{-x1}e^{-x2}
\nonumber\\
&&\rho(x_<)\frac{3 x_>-x_<} {(x_>-x_<)^3} (x_1-x_2)^2(x_3-x_4)^2.
\end{eqnarray}
 In comparison the
total collision rate for s-wave scattering yields
\begin{eqnarray}
\gamma_{s}&=& N(t)
\frac {m\sigma_{s}}{\pi^2 \hbar^3} (\hbar\omega)^2
 {I_s\over \overline T},
\end{eqnarray}
with $I_s=0.5$ is the value of the integral similar to (\ref{ip})
but only with $\rho(x_<)$ in the second line. In the above all
results were given for a spherically symmetric trap with trap
frequency $\omega$. The case of anisotropic traps can be obtained
by substitute $\omega$ with $(\omega_x\omega_y\omega_z)^{1/3}$.

On the other hand, the proposed p-wave collision (inside the
electric field) would not experience such a slow down.
Inside a dc-electric field, the p-wave scattering has the
same non-momentum ($k$) dependence as for the s-wave.
We will now detail the treatment for a single term
$F_F(\vec q,\vec q')=4\pi\,t_{10}^{10}Y_{10}^*(\hat q)Y_{10}(\hat q')$
in the homogeneous case. Using approaches similar to
that lead to Eq. (\ref{pwave}), we obtain
\begin{eqnarray}\label{ptwave}
\rho(\epsilon_4){\partial\over \partial t} n(\epsilon_4,t)
&= &\frac {96m}{h^3}{\pi\over (\hbar\omega)^3} |t_{10}^{10}|^2
\int d \epsilon_{1}\,d \epsilon_{2}\,d \epsilon_{3}\,
\delta(\epsilon_{1}+\epsilon_{2}-\epsilon_{3}-\epsilon_{4})
\nonumber\\
&&{(\sqrt{\epsilon_>}-\sqrt{\epsilon_<})^2}
{(\epsilon_1-\epsilon_2)^2(\epsilon_3-\epsilon_4)^2\over
(\epsilon_>-\epsilon_<)^3}
\int_1^{b}dq {(1+b^2-q^2-b^2/q^2)^{3/2} \over 1+b^2-q^2}
\nonumber\\
&&\Big[n(\epsilon_1)n(\epsilon_2) [1+\eta n(\epsilon_3)][1+\eta
n(\epsilon_4)]- [1+\eta n(\epsilon_1)][1+\eta n(\epsilon_2)]
n(\epsilon_3)n(\epsilon_4)\Big],
\end{eqnarray}
where
$b=(\sqrt{\epsilon_>}+\sqrt{\epsilon_>})
/(\sqrt{\epsilon_>}-\sqrt{\epsilon_>})$.
Although we have not been above to complete the inside integral
analytically, we can immediate see that the energy dependence
for the collision rate is such that a run away evaporation
is possible in the classical limit, i.e.
\begin{eqnarray}
\gamma_p^{\cal E}\sim N(t)
{m |t_{10}^{10}|^2 \over \pi^2\hbar^3}
(\hbar\omega)^2\,{1\over \overline T}.
\end{eqnarray}

We have simulated the evaporation process for the recent experiment
at JILA \cite{jin}. We assume that two hyperfine states were in the
same magnetic trap, and include two types of collisions: the s-wave
collision between the two states and the p-wave collision within
each hyperfine state as discussed above. Inelastic processes are
ignored. We start with $10^7 $ particles in each state at an
initial temperature which is 135 times the initial Fermi
temperature. Then we evaporatively cool the mixture by cutting off
the hot atoms (from both states) with a cut rate which is $1/20$
times the initial collision rate. The atoms rethermalize via p-wave
and s-wave scattering. The p-wave collision rate decreases as the
temperature of the gas decreases whereas the s-wave collision rate
increases due to run-away evaporative cooling down to a temperature
when Fermi-statistics become important and the final state blocking
term $1-n(\epsilon)$ causes the collision rate to decrease. At the
end of the cooling process the temperature is more than 9 times
smaller than the final Fermi temperature and there are less than
$1.3\times 10^6$ particles left in each state. This corresponds to
an increase of phase-space density of more than $10^{12}$. The
final average peak density in the trap is then $\overline
n=4.0\times 10^{12}/$cm$^3$

In Fig. \ref{fig8} we show our results for the numerically computed
collision rates. We note that the p-wave collision rate essentially
dies out at the temperature measured by the JILA experiment
\cite{jin}.
 The sum of the two
scattering rates initially decreases due to the suppressed p-wave
collisions, the total rates eventually increases as the runaway
evaporation sets in for the s-wave scattering. We started the
evaporation at an initial temperature of $\overline T_i=5.23\times
10^5$ and have used the parameters of
$\sigma_s=\sigma_p=10^{-11}$cm$^2$ at $\overline T_0=\overline
T_i/4$ from the JILA results \cite{jin}. Assuming an average trap
frequency of $\omega\approx (2\pi)\,80$ (Hz) this corresponds to an
initial temperature of $T_i=200$ ($\mu$K). The simulation was
started with 500 energy bins and each bin covers an energy interval
of $800\,\hbar\omega$. After the cut-energy is ramped down to 40
energy bins, the energy interval is decreased by a factor 10, a
redistribution of all particle numbers to 500 finer energy bins is
required. This procedure is repeated twice before one finally gets
down to the required low temperature. The accuracy and convergence
were checked by simulations with larger number of energy bins. In
Fig. \ref{fig9} we show the temperature of the evaporated mixture
as well as the corresponding Fermi temperature [computed in terms
of $N(t)$ and $T(t)$]. We note that with appropriate initial
conditions the gas will cross the Fermi degeneracy, at a time
$t_{\rm deg}=3.3\times 10^{-3}\omega^2$ for the parameters used in the
simulation , e.g. 827 (s) for $\omega/2\pi=80$ Hz. We have not
included inelastic loss processes such as three body collisions,
however as long as the inelastic loss rate G is such that
$G\overline n^2<t_{\rm deg}^{-1}$, the cooling process will
successfully reach the degenerate regime.

\section{Conclusion}
In conclusion, we have proposed the possibility of evaporative
cooling of trapped fermionic atoms in an external electric field.
We have derived effective quantum kinetic equations for fermionic
atoms with anisotropic dipole-dipole interactions, and presented
numerical simulations for fermionic evaporative cooling. We have
also discussed sympathetic cooling of fermionic mixtures, related
to a recent experiment \cite{jin}, when two hyperfine states of
$^{40}$K are magnetically trapped. Cooling is principally due to
the predominant s-wave collisions between the two states. We have
observed intrastate p-wave suppression consistent with the
experimental results. In the region where degeneracy becomes
important we have contrasted the qualitative differences in the
evaporative cooling dynamics of bosons and fermions.

This work is supported by the ONR grant 14-97-1-0633 and by the NSF
grants No. PHY-9722410 and 9803180. We want to thank Drs. R. Hulet
and C. Greene for providing the triplet potentials for Li and Rb
respectively. We thank A. Alford for help with the construction of
the triplet potential for K. L.Y. would like to thank many enlightening
discussions with participants of the ITP BEC workshop.

\begin{figure}
\psfig{figure=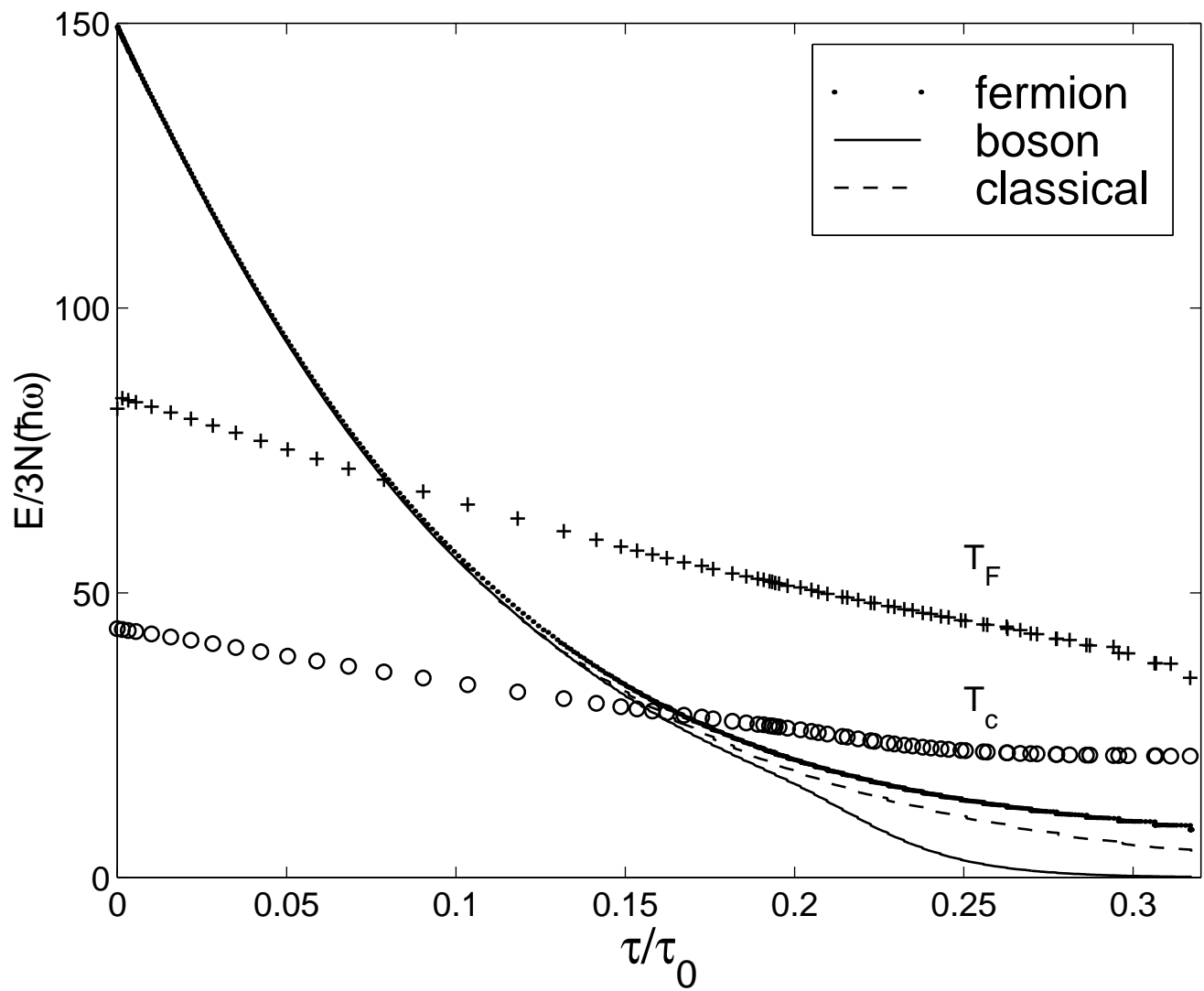}
\vspace{0.5in}
\caption{The same as in Fig. 4, but for the average energy.
}
\label{fig7}
\end{figure}

\begin{figure}
\psfig{figure=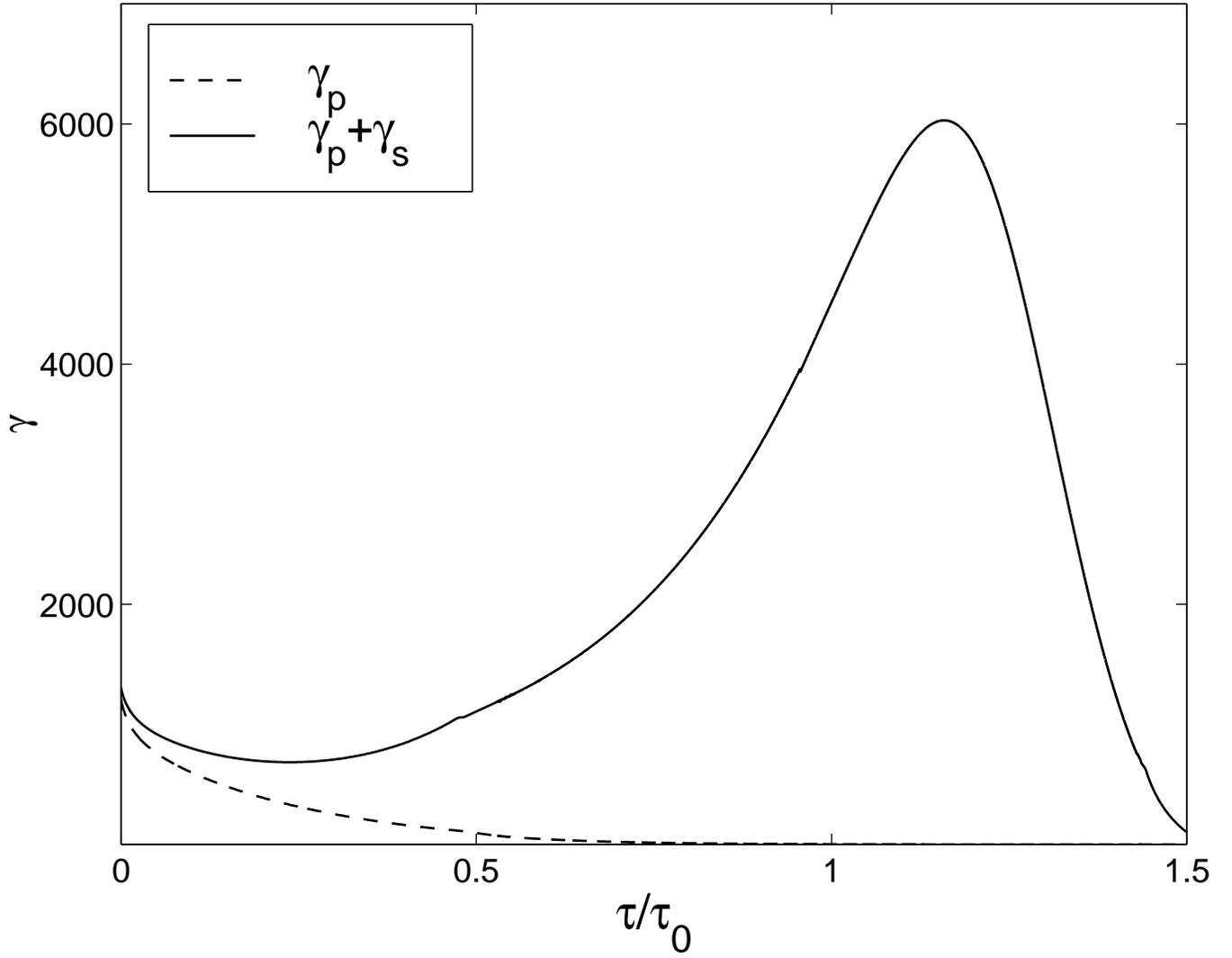}
\vspace{0.5in}
\caption{
The numerically computed collision rates $\gamma_p$ and $\gamma_s$
for the discussed two state evaporation simulation.}
\label{fig8}
\end{figure}

\begin{figure}
\psfig{figure=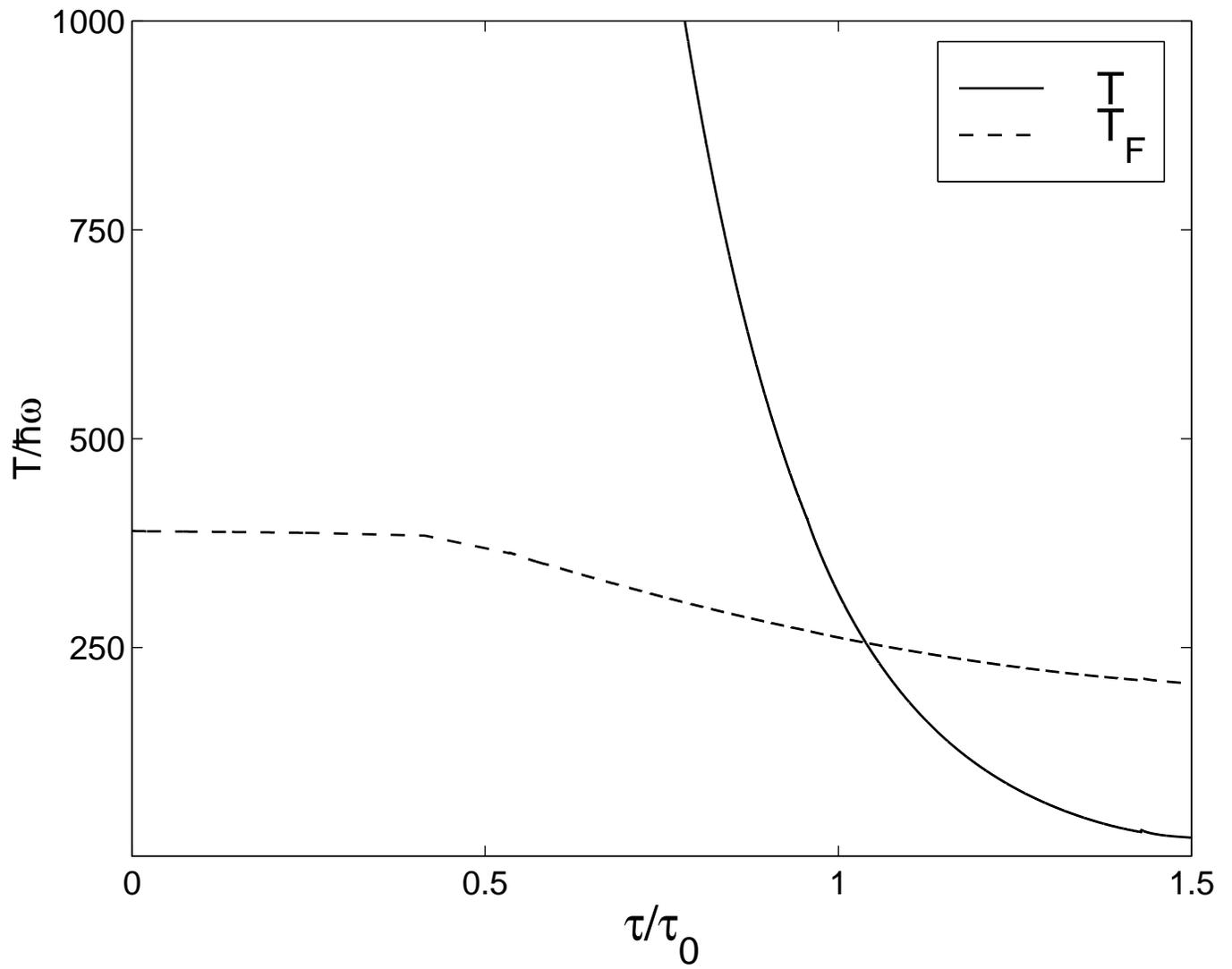}
\vspace{0.5in}
\caption{The temperature and the Fermi temperature for the
evaporated two state Fermi mixture.}
\label{fig9}
\end{figure}

\end{document}